%% file: main.tex
\documentclass[]{spie}

\usepackage{amsmath,amsfonts,amssymb}
\usepackage{graphicx}
\usepackage[colorlinks=true, allcolors=blue]{hyperref}
\usepackage{float}
\usepackage{lineno}

\title{The SPT-3G$+$ receiver design}

\input{athreya_2026_authors}

\authorinfo{Corresponding author: Hrushikesha Athreya, athreya7@uchicago.edu}

\begin{document} 
\maketitle

\begin{abstract}
We present the thermo-mechanical design of the cryostat and camera optics for SPT-3G$+$, a new receiver being developed for the South Pole Telescope (SPT). The receiver consists of 14 detector arrays of 90/150 GHz dichroic polarization-sensitive pixels, totaling 24,080 transition-edge sensor detectors. Each detector array lies at the end of an optics tube, each approximately 240 mm in diameter and 772 mm in length, which are arranged in a hexagonal close-packed configuration to achieve a 4 degree diameter field of view. Each optics tube contains four anti-reflection coated lenses fabricated from different materials (alumina, silicon, and nylon) that are designed to also provide infrared filtering that reduces the radiative loading on the cryogenic stages. The optics and detectors are cooled by a combination of a pulse tube cooler for the 40\,K and 4\,K stages, and a dilution refrigerator for the 1\,K and 100\,mK stages. Thermal modeling predicts the heat load to be less than 26\,W and 1\,W for the 40\,K and 4\,K stages, respectively. The 1,550 kg cryostat has a 1.1 meter diameter at the vacuum window, which is located near the telescope Gregorian focus, and 1.75 meters in height and length. Fabrication of the cryostat will begin in 2026, with installation on the SPT scheduled for the 2028-29 austral summer, ahead of the 2029 winter observing season.  
\end{abstract}


\section{INTRODUCTION}
\label{sec:intro}  

The South Pole Telescope (SPT) is a 10-meter diameter, submillimeter-wavelength telescope with arcminute resolution located at the Amundsen-Scott South Pole Station \cite{carlstrom2011}. The South Pole offers a stable, dry, high, and radio-quiet environment, making it an ideal site for observations of the cosmic microwave background (CMB) \cite{bussmann05}. 

Inflation posits a field (or fields) that drives a brief period of rapid expansion during the first fraction of a second of the universe. During this period of expansion, quantum fluctuations generate a background of inflationary gravitational waves (IGWs), which induce a B-mode polarization pattern on the CMB.  However, current B-mode measurements need to be able to disentangle foreground signals, such as B modes generated from the gravitational lensing of CMB E modes and Galactic dust, from the faint IGW signal \cite{zebrowski2025}.

The South Pole Observatory (SPO) is a coordinated observing and analysis program between the BICEP and SPT collaborations that aims to constrain the tensor-to-scalar ratio $r$ --- the parametrization of the IGW signal --- to $\sigma(r)= 0.001$. The BICEP experimental program consists of a series of receivers measuring the CMB at degree angular scales in order to test predictions of inflation and constrain $r$ \cite{moncelsi20}. The current best constraint on $r$ comes from BICEP, with $\sigma(r) = 0.009$ \cite{bicep2keck21b}. The next-generation camera for the SPT, SPT-3G$+$, is optimized to measure CMB polarization at arcminute angular scales with sufficient sensitivity to remove the lensed B-mode signal from the BICEP data to achieve $\sigma(r) = 0.001$.  SPT-3G$+$ will also constrain cosmology through measurements of the CMB lensing spectrum and galaxy clusters \textcolor{blue}{Natoli et al., these proceedings}. 

SPT-3G$+$ has nearly an order of magnitude higher mapping speed compared to the currently installed SPT-3G camera \cite{sobrin22}. This increase in sensitivity is achieved through multiple factors, including improvements in cryogenics, optics, detectors, and readout electronics. In this manuscript, we describe the design of the cryostat and optics tubes for SPT-3G$+$, highlighting the changes relative to SPT-3G that enable this improvement in mapping speed. For more information on SPT-3G$+$'s science goals and other subsystems, see \textcolor{blue}{Natoli et al., these proceedings}.

\section{CRYOSTAT OVERVIEW}
\label{sec:cryostat}

   \begin{figure} [H]
   \begin{center}
   \begin{tabular}{c}  
   \includegraphics[trim=1.3cm 4cm 0cm 3cm, clip, width=1.0\textwidth]{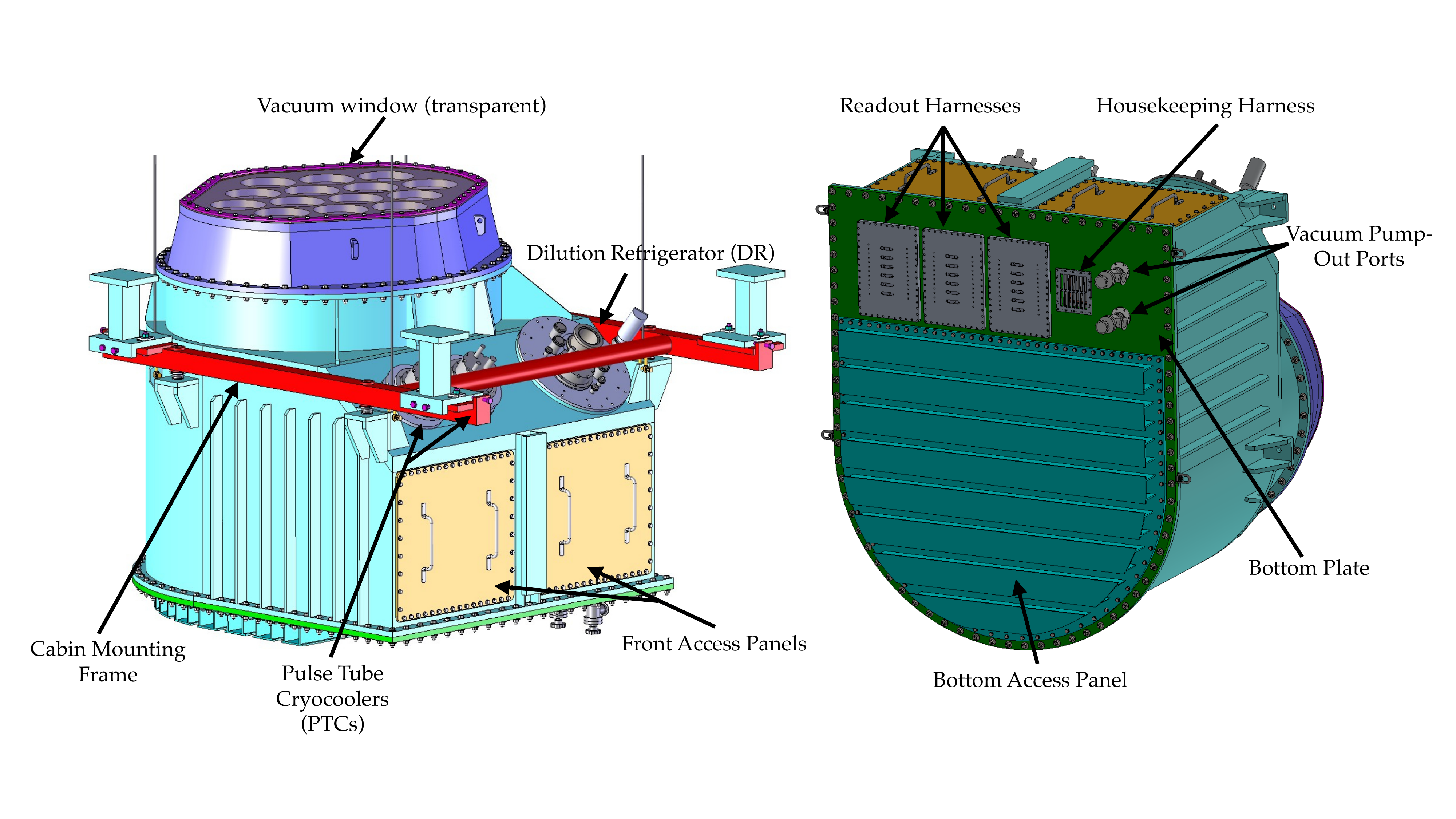}
   \end{tabular}
   \end{center}
   \caption[example] 
   { \label{fig:cryostat} 
A side-view (left) and bottom-view (right) of the cryostat showing the cabin mounting frame, refrigerators, and wire harnesses installed. The vacuum window is made transparent in this figure to show the vacuum plate apertures for the optics tubes.}
   \end{figure} 

The SPT-3G$+$ receiver is a cryogenic vacuum vessel with three shells, a 300\,K vacuum shell, a 40\,K radiation shield, and a 4\,K radiation shield. The 1,550 kg cryostat occupies a cubic space of width 1.8 meters, length 2.0 meters, and height 1.7 meters. The cryostat will house 14 optics tubes (OTs), which additionally have 1\,K and 100\,mK stages, as described in Section~\ref{sec:opticstubes}. The number of OTs was chosen to be the maximum that can be accommodated within the SPT cabin while maintaining high optical quality and negligible vignetting from the warm optics across the entire field of view. Each shell of the cryostat will have front and bottom access panels, as shown in Figure \ref{fig:cryostat}; the front access panels provide access to the readout and housekeeping harnesses, detailed in Section~\ref{sec:URH}, while the bottom access panel provides access to the OTs and heat strapping connections. The cryostat will deploy with one BlueFors SD250 Dilution Refrigerator (DR)\footnote{https://bluefors.com/products/dilution-refrigerator-measurement-systems/sd-dilution-refrigerator-measurement-system/} and one PT420 Pulse Tube Cryocooler (PTC)\footnote{https://bluefors.com/products/pulse-tube-cryocoolers/pt420-pulse-tube-cryocooler/}, with room to install a second PTC should in-lab testing indicate the need for additional cooling capacity. 
The Bluefors DR will cool the detector arrays to 100\,mK, $\sim$3 times colder than the SPT-3G detectors \cite{sobrin22}, which is one factor that increases the relative mapping speed of SPT-3G$+$.  In addition, the continuous DR is able to maintain steady base temperatures without cryogenic cycling, further increasing the relative observing efficiency.  

\subsection{300\,K Vacuum Shell}
\label{sec:300K}

\begin{figure} [ht]
   \begin{center}
   \begin{tabular}{c}  
   \includegraphics[height=8cm]{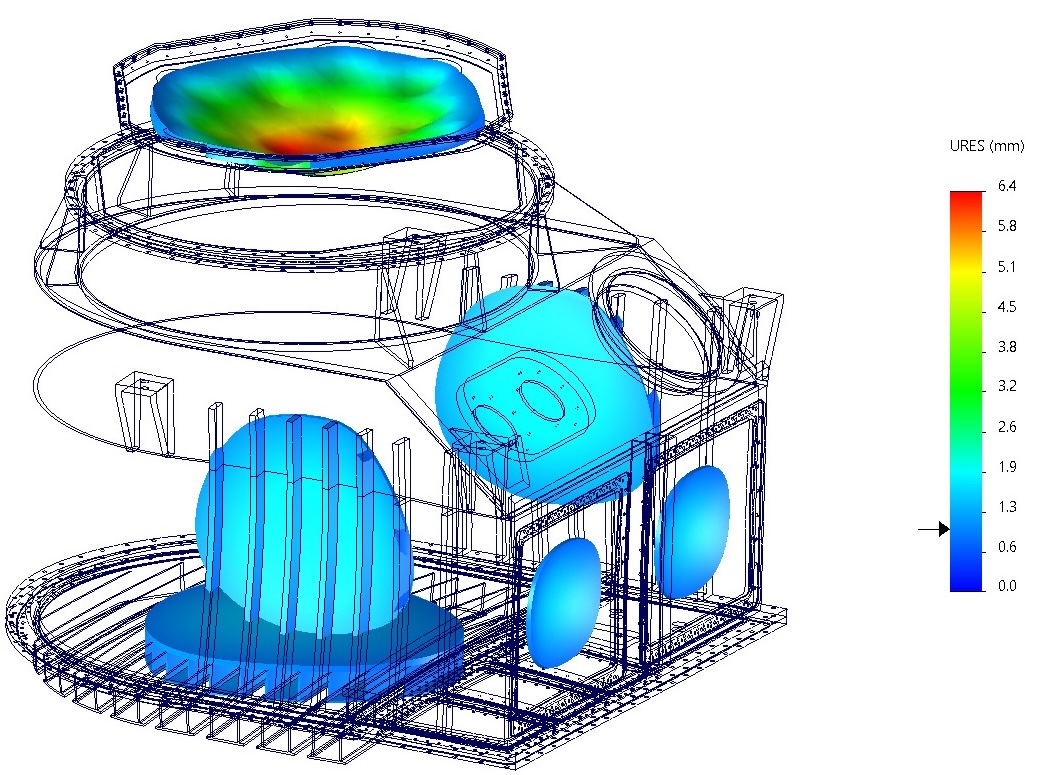}
   \end{tabular}
   \end{center}
   \caption[example] 
   { \label{fig:vacuum_shell_FEA} 
Finite element analysis (FEA) of the vacuum deflection experienced by the 300\,K vacuum shell under 1 atmosphere of pressure (0.1 MPa). Shown here are all the locations of larger than 1 mm deflection. Deformation scale is 31.  The maximum deflection is $\sim$6.5 mm at the center of the vacuum window at the top of the cryostat.}
\end{figure} 
The 300\,K vacuum shell of the cryostat is made from 5083 aluminum, welded together with thicknesses ranging from 0.5 to 2 inches to withstand vacuum deflection at various positions, as shown in Figure~\ref{fig:vacuum_shell_FEA}. Our finite element analysis (FEA) predicts a maximum deflection of roughly 6.5 mm under vacuum, which still leaves a $\sim$10 mm gap before the first element in the OT. The deflection does not significantly impact the optical design, since the window is simply an aperture in the optical system and not connected to any refracting optical components, for more details, see Section \ref{sec:opticstubes}. The cryostat will attach to the optics bench in the SPT cabin via four mounting locations around the perimeter of the vacuum shell, seen in Figure \ref{fig:cryostat}. The cryostat is designed such that the cryocoolers will be aligned with gravity when the telescope observes at 40-degrees elevation, ensuring high cooling efficiency during typical observations. The vacuum window of the cryostat is a 6 mm thick sheet of high-density polyethylene (HDPE) that covers all 14 apertures, and is supported by the aluminum vacuum shell below the window. Immediately below the window is an assembly of twelve 1/8~in Zotefoam\footnote{https://www.zotefoams.com/who-we-are/3-stage-process/} infrared filters that will be attached with a stycast adhesive to six 1/16~in aluminum spacers, one between every other sheet, similar to the design used in the SPT-3G cryostat \cite{sobrin22}. Vacuum pump-out ports and the wire harnesses are located on the bottom of the shell for easier integration and access. The bottom plate of the shell is removable to install the radiation shields, which are described in Section~\ref{sec:rad_shields}.    

\subsection{Radiation Shields}
\label{sec:rad_shields}

\begin{figure} [H]
   \begin{center}
   \begin{tabular}{c}  
   \includegraphics[trim=1cm 0cm 1cm 0cm, clip, width=1.0\textwidth]{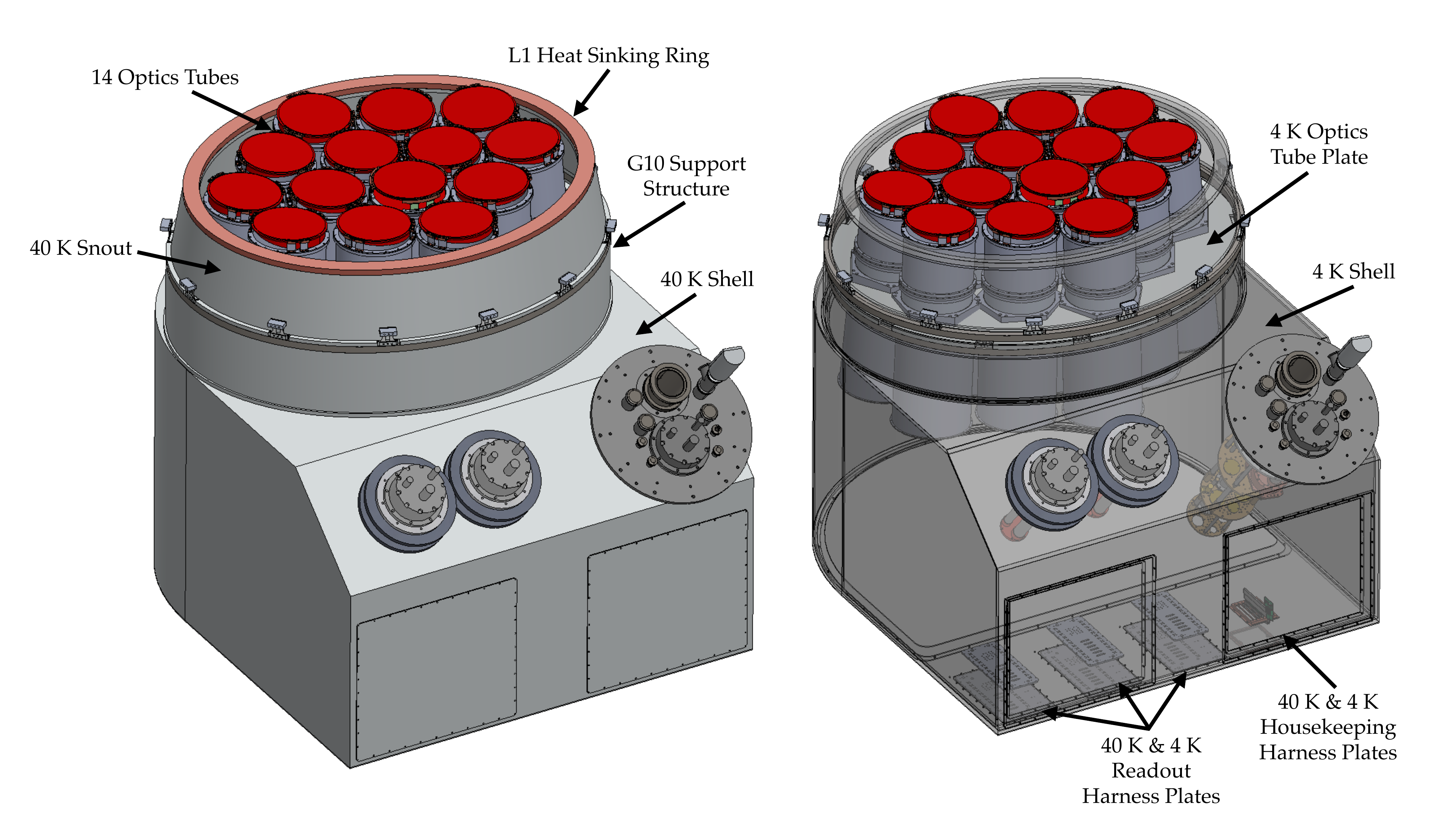}
   \end{tabular}
   \end{center}
   \caption[example] 
   { \label{fig:rad_shields} 
A rendering of the 40\,K and 4\,K radiation shields.  (Left) The exterior of the 40\,K radiation shields can be seen, including the alumina lenses at the top of each OT and the G10 support structure that supports the shield from the 300\,K vacuum shell.  (Right) The 40\,K and 4\,K radiation shields are made transparent, to better see components inside the 4\,K volume, including the radiation shield, bottom of the OTs, and various wire harnesses interfaces on the bottom of the cryostat.
}
\end{figure} 

\begin{figure} [ht]
   \begin{center}
   \begin{tabular}{c} 
   \includegraphics[trim=1cm 1.5cm 1.5cm 4.2cm, clip, width=1.0\textwidth]{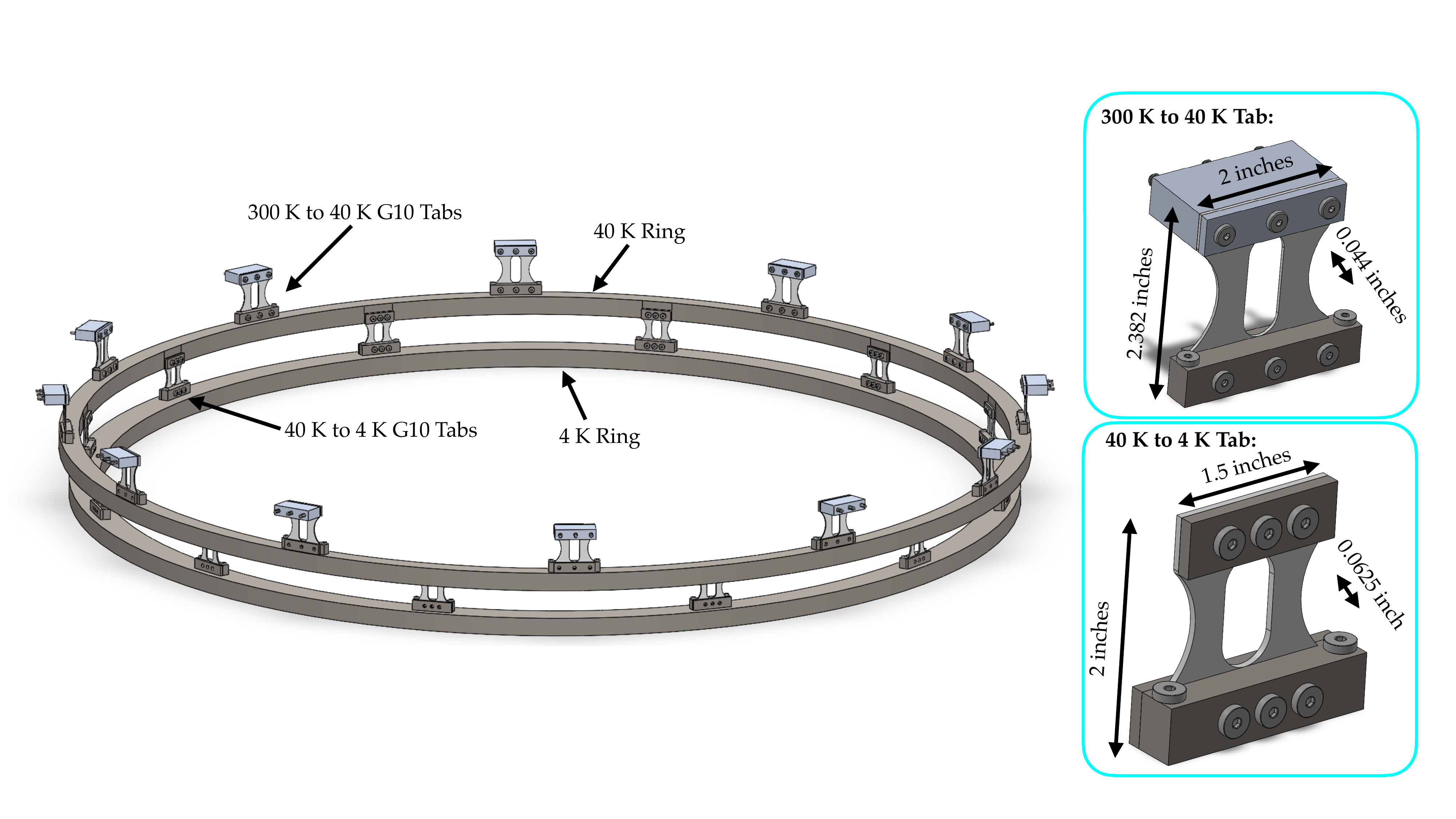}
   \end{tabular}
   \end{center}
   \caption[example] 
   { \label{fig:G10} 
(Left) The design of the G10 support structure that mechanically supports, and physically references, the radiation shields and OTs from the vacuum shell.  The dimensions of the 300\,K to 40\,K G10 tab (right-top) and the 40\,K to 4\,K G10 tab (right-bottom).}
\end{figure} 

The 40\,K and 4\,K radiation shields are shown in Figure \ref{fig:rad_shields}.  The shields are primarily made from 1100 aluminum, which is a lightweight alloy with relatively high thermal conductivity at cryogenic temperatures. Each shield generally has two components that come together on the top and bottom faces of the G10 support structure, which is also shown in more detail in Figure \ref{fig:G10}. A snout-shaped radiation shield and heat-sinking ring are attached at the top of the 40\,K ring in the G10 support structure. These help cool the L1 lenses, further described in Section \ref{sec:opticstubes}. Preliminary thermal FEAs of the radiation shields show small thermal gradients across the shells. At the top of the 4\,K ring attaches a 4\,K optics tube plate, where the OTs are all attached, and is the main mechanical reference that defines the relative positions of the OTs. At the bottom of the 40\,K and 4\,K rings in the G10 support structure are attached nested rectangular-shaped radiation shields at each temperature.

The 40\,K and 4\,K rings of the G10 support structure are made from 316 stainless steel. The structure consists of 12 G10 tabs along the 40\,K ring's elliptical perimeter providing a thermal break from the 300\,K to 40\,K and 12 similar tabs along the 4\,K ring's perimeter, providing a thermal break from 40\,K to 4\,K. The G10 structure design was optimized for mechanical strength, while minimizing thermal conductivity, and allowing for flexibility due to differential thermal contraction from either end of the tabs. Several rounds of finite element analyses (FEAs) that included the effects of differential thermal contraction and a changing gravity vector during telescope operations led to the ``hourglass" shape configuration shown in Figure \ref{fig:G10}. While the shape of the two tabs is similar, their dimensions are slightly different. The 300\,K to 40\,K tabs screw directly into the vacuum shell on the 300\,K side. The metal feet that clamp the G10 tab on that end are made from 6061 aluminum, while the metal feet that clamp the 40\,K end are made from 316 stainless steel. The 40\,K to 4\,K tabs screw directly into the 40\,K ring of the structure and both ends' metal feet are made from 316 stainless steel.  

\subsection{Cryogenic Readout}
\label{sec:URH}

\begin{figure} [ht]
   \begin{center}
   \begin{tabular}{c}  
   \includegraphics[trim=5cm 4.5cm 7cm 2cm, clip, width=1.0\textwidth]{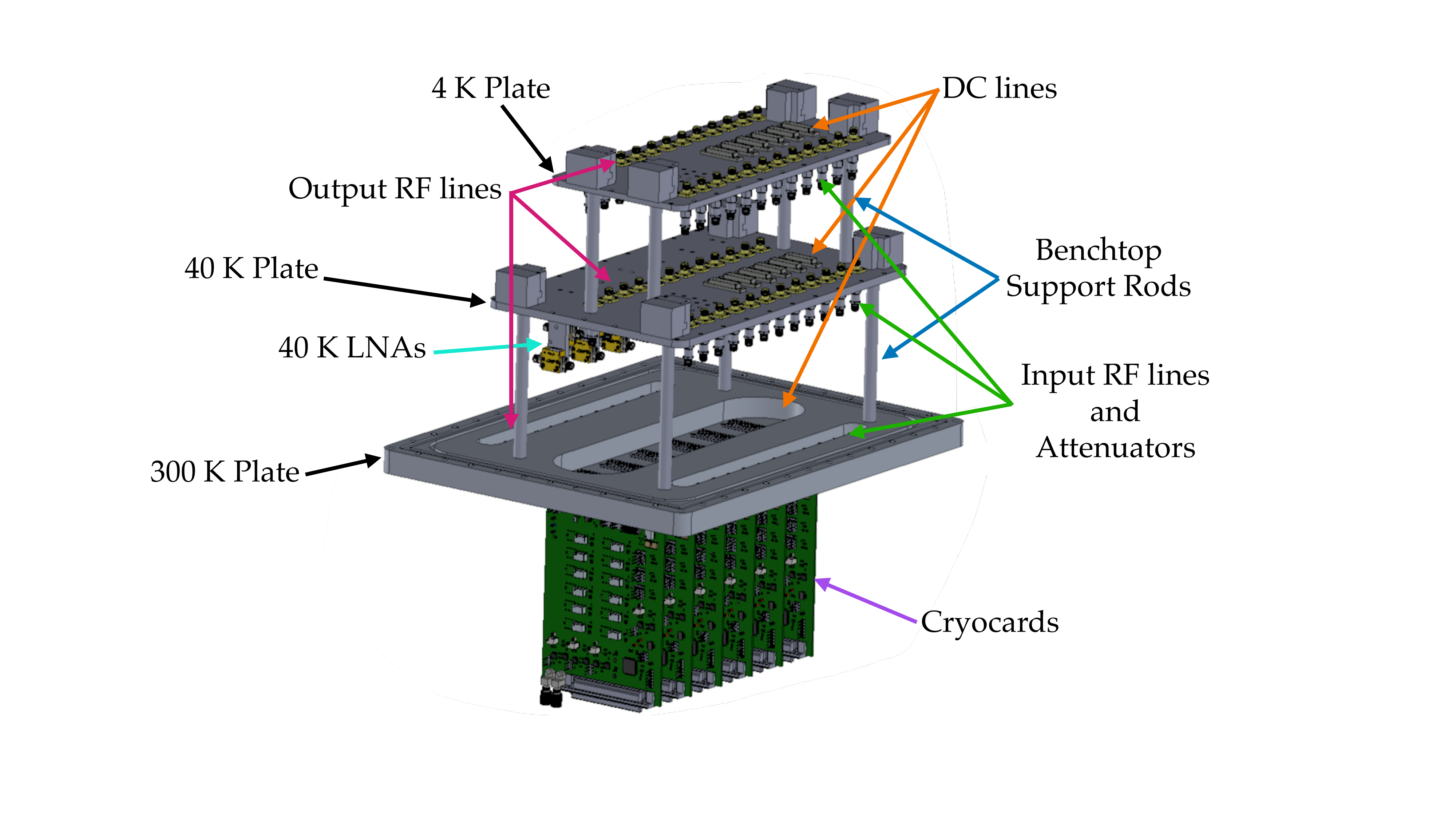}
   \end{tabular}
   \end{center}
   \caption[readoutharness] 
   { \label{fig:readoutharness} 
A rendering showing the design of the readout wire harness. The purpose of the wire harness is to connect the detector readout chain from 300\,K to 4\,K, while providing heat sinking on the 40\,K and 4\,K stages.  The detector readout chain connected to the wire harness includes 40\,K LNAs, coaxial cables, RF attenuators, and DC wiring, though we note that the coaxial cables and DC wiring between stages are not shown in the image.}
   \end{figure} 
   
The cryogenic detector readout is designed to minimally contribute to the overall instrument noise, while minimizing the heat load on the different cryogenic stages. SPT-3G$+$ uses microwave SQUID multiplexing ($\mu$Mux), a frequency-domain approach that couples each detector to a superconducting resonator. Nearly 1000 detectors are read out on a single RF line, roughly an order of magnitude more than SPT-3G \cite{montgomery22}. SLAC Microresonator Radio Frequency (SMuRF) electronics are used to operate the $\mu$Mux \cite{henderson18}. 

Each detector wafer requires two radio frequency (RF) coaxial input lines, two RF coaxial output lines, and at least 28 direct current (DC) lines. The RF lines carry signals between the multiplexer chips that are coupled to the detectors and SMuRF electronics. We use a series of discrete attenuators at each temperature stage to heat sink the coaxial cabling, as well as to control tone powers at the resonators. DC blocks will also be used to provide thermal breaks at specific stages. We use a two-stage cryogenic low-noise amplifier (LNA) system on the RF output line with a low gain HEMT LNA\footnote{Low Noise Factory LNF-LNC4\_8F\_LG} sunk to 4\,K, and another ASU LNA at 40\,K. We also have an LNA\footnote{Mini-Circuits ZX60-83LN12+} located on the outside of the cryostat to optimize signal power input into the SMuRF electronics.

The DC lines to the detector wafer provide bias power for the TES bolometer detectors as well as for flux ramp modulation for the rf-SQUIDs. We also account for additional DC lines that power the cryogenic amplifiers, but these will terminate at 4\,K or 40\,K at their respective amplifier. Thus, in total, for 14 detector wafers, we require 28 RF input and output lines, and over 900 DC lines to traverse through the different temperature stages without exceeding the total cooling budget available for that stage.

To achieve this goal, we designed the cryogenic readout as two main subsystems. The first is the readout wire harness that traverses from the outside of the cryostat at 300\,K down to 4\,K, with an intermediate 40\,K stage that houses the 40\,K LNAs. We use hermetic SMA bulkheads for the RF lines and hermetic 50-pin D-subminiature (D-sub) connectors for the DC lines on the 300\,K stage. The TES bias lines as well as bias for cryogenic amplifiers are controlled by a PCB with bias resistors and filters, which we call a cryocard, that is mounted directly to each D-sub connector on the outside of the cryostat. For the 40\,K and 4\,K stages, however, we plan to use regular SMA bulkheads and 51-pin Micro D-subminiature (MDM) connectors to save space. The power for the 40\,K LNAs will be directly branched out from the 300\,K 50-pin hermetic D-sub. The SPT-3G$+$ cryostat will use three identical readout harnesses, shown in Figure \ref{fig:readoutharness}, each of which can support up to 6 detector wafers, designed so that there are a sufficient number of spare channels to reduce the frequency of having to remove/reinstall the readout harness to replace any components.

The second subsystem is the OT readout assembly, which includes the 4\,K LNA as well as all sub-4\,K readout cabling and components. Each OT will get its own readout assembly that will hang off the back of the detector modules and bring signals from 4\,K to 100\,mK through an intermediate 1\,K stage. Each OT readout assembly connects to one of the readout harnesses via isothermal lines that will be routed along specially designed raceways to ensure that there are no interferences with heat strapping that could cause a thermal short. For each OT readout assembly, we will use a PCB to breakout the DC lines coming from the readout harness to power the 4\,K LNAs and for the detector wafer. The 4\,K LNAs sit on dedicated ``diving boards" mounted around the 4\,K shell near each OT to minimize signal loss preamplification as well as providing thermal pathways for heatsinking. Although not currently included in the readout design, we are keeping the option to potentially include a cryogenic isolator\footnote{Low Noise Factory LNF-ISC4\_8A} at 1\,K to block input reflections from the 4\,K LNA into the detector module. The decision on whether or not we use this isolator is pending empirical testing in lab.

To minimize heat flow to the colder temperature stages, we will be using a selection of stainless steel/cupro-nickel coaxial cabling for the RF lines and 36 AWG phosphor-bronze for the DC wiring due to their low thermal conductivity. To further minimize signal loss, we will be using superconducting niobium-titanium for the output RF coaxial cabling from the detector wafer to the 4\,K LNA and for all DC lines at sub-4\,K stages.

We will attach thermometry -- diodes and negative temperature coefficient resistance temperature detectors (RTDs) -- around various locations in the cryostat to be read out via a housekeeping harness in order to monitor the cryogenic performance of the cryostat. For this harness, we will be adapting one of the wiring harnesses from SPT-3G to bring the required circuitry from 300\,K to 4\,K, which can then be routed to each thermometer.

\section{OPTICS TUBES}
\label{sec:opticstubes}

   \begin{figure} [ht]
   \begin{center}
   \begin{tabular}{c}  
   \includegraphics[trim=4cm 0cm 4cm 0cm, clip, width=1.0\textwidth]{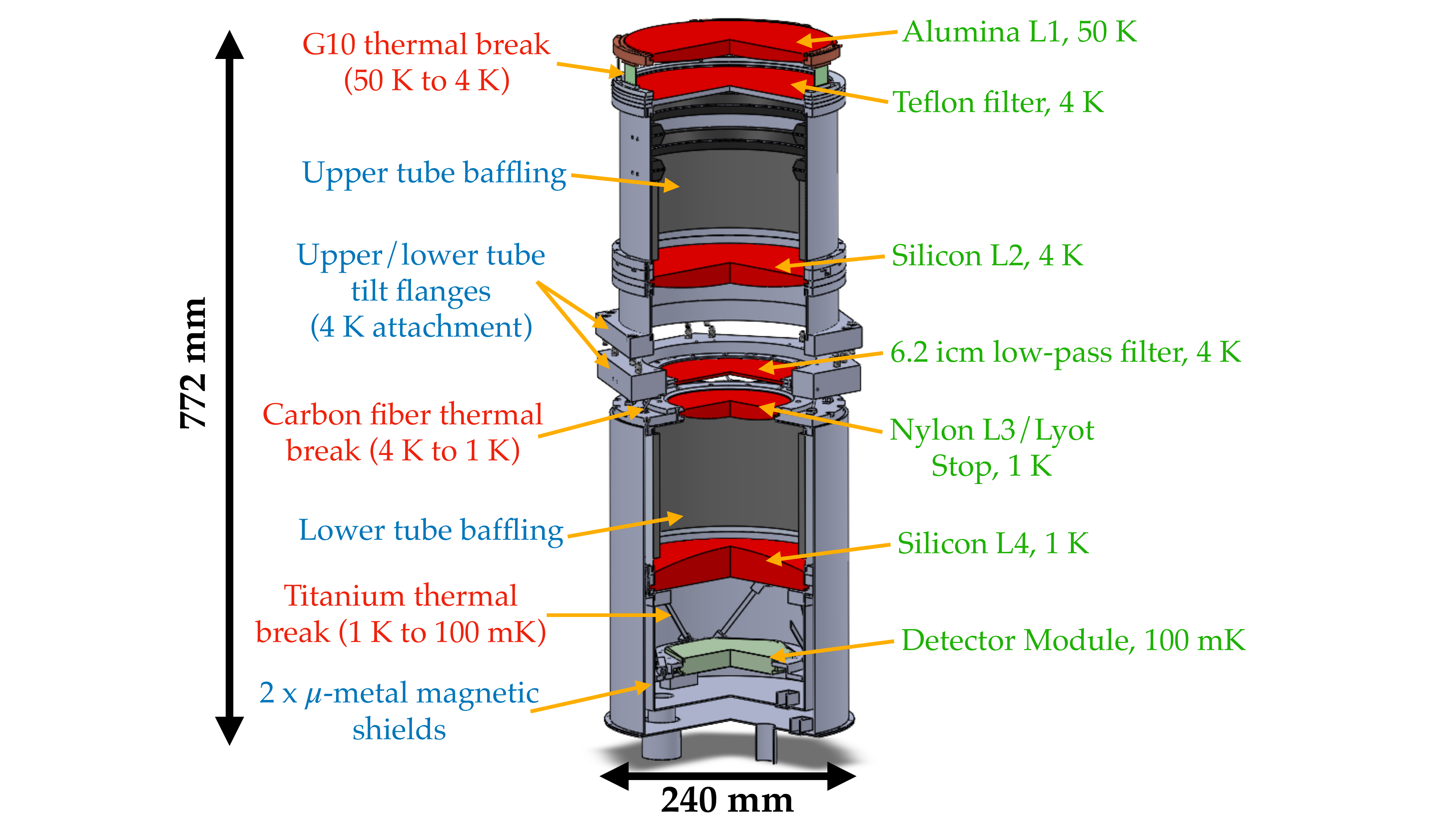}
   \end{tabular}
   \end{center}
   \caption[example] 
   { \label{fig:optics_tube} 
The design of an optics tube, with a cutaway showing the different elements and stages. For simplicity, we show the central tube closest to the telescope boresight, however the other tubes have a similar design but with a several-degree relative tilt that follows the curved Gregorian focus.  Green labels highlight optical elements, red labels highlight thermal breaks, and blue labels highlight structural components.  The design enables the different temperature optical components to be mechanically referenced to each other, to ensure accurate optical alignment.}
   \end{figure} 

The SPT-3G$+$ cryostat will house 14 optics tubes each approximately 240 mm in diameter and housing a detector wafer containing 430 pixels, as shown in Figure \ref{fig:optics_tube}. Although each OT contains the same lens and filter materials, the thicknesses and dimensions of the optical elements vary from tube to tube based on the optical design. Each OT contains four meta-material anti-reflection coated lenses \cite{golec20} and two filters: a Teflon filter placed between the first and second lenses (L1 and L2), and a 6.2 icm low-pass (metal-mesh) filter \cite{ade2006} placed before the third lens (L3). Lenses are secured with axial and radial spring gasketing to account for differential thermal contraction between the lens and the mount. Alignment notches are also incorporated into every tube component to facilitate repeatable and accurate integration and assembly.

Before reaching L1, incident light passes through a stack of Zotefoam infrared blockers\cite{sobrin22}. L1 is fabricated from alumina, which serves simultaneously as an infrared filter and a refractive lens. A copper ring on the 40\,K shell, labeled as ``L1 Heat Sinking Ring" in Figure~\ref{fig:rad_shields}, connects to all 14 L1 elements via a network of heat straps, providing a common heat sink. While the first radiation shield of the cryostat will be at 40\,K, our thermal FEA predicts that, based on the thermal loading at L1 and heat strap connections to the L1 heat sinking ring, the L1 lenses will be closer to 50\,K. L2 and L4 are both fabricated from silicon, which has limited contraction at cryogenic temperatures and limited re-emission within our observing bands. The L2 and L4 lenses are thermally heat sunk to the 4\,K and 1\,K stages, respectively.

Each OT is divided into a top half, containing optical elements from L1 through L2, and a bottom half, containing elements from the metal-mesh filter down to the detector module. Both halves mount to opposite sides of the 4\,K optics tube plate, as shown in Figure~\ref{fig:rad_shields}, and mechanically referenced to each other via alignment pins. The top half is thermally coupled to the 4\,K optics tube plate, providing a heat sink for L2. Since L1 operates at 50\,K, it is mounted to the top tube through G10 tabs, which accommodate differential thermal contraction and provide a thermal break between the two temperature stages.

On the bottom tube side, the metal-mesh filter mounts to the flange that attaches to the 4\,K optics tube plate. Carbon fiber struts arranged in a ring around this flange support the bottom tube and provide a thermal break between the 4\,K optics tube plate and the 1\,K bottom tube. Immediately inside of the carbon fiber struts sits the L3 lens, fabricated from nylon and located at the Lyot stop. Nylon serves a dual role: it acts as an infrared filter that blocks high-frequency radiation from the 100\,mK stage \cite{halpern86}, and as a refractive lens element. In a similar fashion to the carbon fiber struts, titanium struts (Ti 15-3-3-3) around the base of the L4 lens support the detector module and provide a thermal break between the L4 lens at 1\,K and the detector module at 100\,mK.  Titanium becomes superconducting below 1\,K, making it an effective thermal isolator at these temperatures \cite{kellaris14}.

In order to mitigate environmental effects from Earth's magnetic field on detector elements \cite{huber22}, the 1\,K and 100\,mK portions of the OTs are enclosed in two nested layers of $\mu$-metal magnetic shielding. Both shields incorporate ``chimneys" to allow a cold finger to protrude for 100\,mK and 1\,K heat sinking. Heat straps attached to the side of the bottom tube provide 1\,K heat sinking to the bottom tube assembly. Flexible oxygen-free high-conductivity (OFHC) copper braided heat straps will connect the DR to these heat sinking locations on the OTs. 

\section{Cryogenics Overview}
\label{sec:cryogenics}

The predicted thermal loading for each cryogenic stage is summarized in Table~\ref{tab:Thermal_Loading}. The support structure row highlights the power transferred conductively through the G10 tabs that hold up the inner 40\,K and 4\,K shells described in Section~\ref{sec:rad_shields}, and the carbon fiber and titanium supports for the 1\,K and 100\,mK sections, respectively, described in Section~\ref{sec:opticstubes}. The radiation row describes the radiative power loading on each stage from the surrounding radiation shields. The window row describes the power radiated from either optical or infrared filtering elements onto each stage, assuming the OT design described in Section~\ref{sec:opticstubes}. The readout and housekeeping loadings are powers sunk to the radiation shields from the wiring in those systems, as described in Section~\ref{sec:URH}.     
\begin{table}[ht]
\caption{The predicted thermal loading for each stage of the SPT-3G$+$ cryostat.  The predicted loading is at least a factor of two below the available cooling power on each stage.} 
\label{tab:Thermal_Loading}
\begin{center}       
\begin{tabular}{|l|l|l|l|l|}
\hline
\rule[-1ex]{0pt}{3.5ex}   & \textbf{40\,K Stage} & \textbf{4\,K Stage} & \textbf{1\,K Stage} & \textbf{100\,mK Stage}  \\
\rule[-1ex]{0pt}{3.5ex} \textbf{Source} & \textbf{(W)} & \textbf{(W)} & \textbf{(mW)} & \textbf{($\mu$W)}  \\
\hline
\rule[-1ex]{0pt}{3.5ex}  Support Structure & 1.5 & 0.38 & 5.2 & 21.1   \\
\hline
\rule[-1ex]{0pt}{3.5ex}  Radiation & 9.5 & 0.09 & $<$0.1 & 0.9  \\
\hline
\rule[-1ex]{0pt}{3.5ex}  Window & 13.0 & 0.16 & $<$0.1 & 0.1  \\
\hline
\rule[-1ex]{0pt}{3.5ex}  Readout & 1.7 & 0.09 & 0.1 & 7.5  \\
\hline 
\rule[-1ex]{0pt}{3.5ex}  Housekeeping & 0.06 & $<$0.01 & $<$0.1 & 0.2  \\
\hline 
\hline 
\rule[-1ex]{0pt}{3.5ex}  Total loading & 25.8 & 0.72 & 5.3 & 29.8  \\
\hline
\rule[-1ex]{0pt}{3.5ex}  Available power & 55 & 2 & 15 & 250  \\
\hline 
\end{tabular}
\end{center}
\end{table}

As also noted in Section \ref{sec:cryostat}, the cooling capacity for the 40\,K and 4\,K stages of the cryostat comes from a PTC, Cryomech model number PT420.  The total loading on the 40\,K stage is 25.8\,W, roughly a factor of 2 under the 55\,W cooling power at 45\,K nominally specified from the first stage of the PT420. The total loading on the 4\,K stage is 0.71\,W, well within the 2\,W of cooling power at 4.2\,K nominally specified from the second stage of the PT420.  

At the 1\,K and 100\,mK stages, the cooling capacity comes from the DR, Bluefors model number SD250, which provides approximately 15\,mW of cooling power from the still at the 1\,K stage, and 250\,$\mu$W from the mixing chamber at the 100\,mK stage. The total predicted load on the 1\,K stage is approximately 5.26\,mW, well within the available cooling power. At the 100\,mK stage, the total predicted load is approximately 29.8\,$\mu$W, again comfortably within the DR's 250\,$\mu$W cooling capacity, leaving meaningful margin for excess thermal gradients. It is important to note that we are not using the PT415 that comes equipped within the DR to sink the radiation shields. 

\section{SUMMARY AND CURRENT STATUS}
We have described the design of the SPT-3G$+$ receiver, highlighting the vacuum and radiation shields' structure, cryogenic readout system, optics tube layout, and thermal budget. This design, along with the detector packing density highlighted in \textcolor{blue}{Natoli et al.}, will enable  SPT-3G$+$ to have a mapping speed nearly an order of magnitude larger compared to the current SPT-3G receiver. 
SPT-3G$+$'s measurements will enable new constraints on cosmology and inflation, with the lensing map used to delens foreground B modes of the combined SPO data set with BICEP to achieve a measurement of $\sigma(r) = 0.001$. 

The cryostat shells are expected to arrive by the end of 2026 and optics tube component testing is scheduled to begin at the start of 2027. Optics tube and readout harness integration will occur in 2027 and the receiver will be installed onto SPT during the 2028-29 austral summer, ahead of the 2029 winter observing season. 

\acknowledgments      
\input{acknowledgements}

\bibliography{3G+_athreya} 

\bibliographystyle{spiebib} 

\end{document}

%% file: athreya_2026_authors.tex
\author[a,b,c]{H.~Athreya}
\author[d,e]{Z.~Ahmed}
\author[f]{J.~E.~Austermann}
\author[f]{K.~Bae}
\author[g,c]{A.~Bapat}
\author[h]{D.~R.~Barron}
\author[i]{P.~S.~Barry}
\author[c,b,a]{A.~N.~Bender}
\author[j,b,a]{B.~A.~Benson}
\author[c,b,a]{L.~E.~Bleem}
\author[b,k,c,a,l]{J.~E.~Carlstrom}
\author[c]{T.~W.~Cecil}
\author[b,c,a]{C.~L.~Chang}
\author[e]{S.~Cisneros}
\author[m]{A.~Coerver}
\author[c]{J.~Cornelison}
\author[a]{R.~Datta}
\author[a,b]{K.~R.~Dibert}
\author[h]{W.~Dominguez}
\author[f]{S.~M.~Duff}
\author[k,b]{K.~Fichman}
\author[n,o]{J.~P.~Filippini}
\author[p]{L.~Gades}
\author[q,b]{P.~A.~Gallardo}
\author[r]{S.~Galli}
\author[s,t]{N.~W.~Halverson}
\author[o]{Q.~Hao}
\author[e]{S.~Henderson}
\author[e]{R.~Herbst}
\author[m]{W.~L.~Holzapfel}
\author[k,b]{A.~Hryciuk}
\author[f]{J.~Hubmayr}
\author[f,t]{D.~Jones}
\author[k]{V.~Kabra}
\author[u]{K.~S.~Karkare}
\author[v]{C.~King}
\author[f,t]{M.~A.~Koc}
\author[b,a,k]{A.~M.~Kofman}
\author[f,w]{R.~A.~Lew}
\author[f]{M.~J.~Link}
\author[f]{T.~J.~Lucas}
\author[k]{A.~Mangu}
\author[a,b]{E.~S.~Martsen}
\author[b,k,a]{J.~J.~McMahon}
\author[x]{J.~Montgomery}
\author[v]{J.~M.~Nagy}
\author[a,b]{T.~Natoli}
\author[j]{H.~Nguyen}
\author[y,z]{V.~Novosad}
\author[aa]{S.~Padin}
\author[e]{T.~Pinsonneault-Marotte}
\author[bb,cc]{S.~Raghunathan}
\author[a,b]{A.~S.~Rahlin}
\author[dd]{C.~L.~Reichardt}
\author[e]{L.~Ruckman}
\author[v]{J.~E.~Ruhl}
\author[v]{M.~S.~Sarwar}
\author[j,a]{S.~Simon}
\author[f]{R.~Singh}
\author[ee,j]{J.~A.~Sobrin}
\author[ff]{A.~A.~Stark}
\author[v]{Q.~Taylor}
\author[i]{C.~Tucker}
\author[f]{J.~Ullom}
\author[f]{J.~Van Lanen}
\author[o,n,cc]{J.~D.~Vieira}
\author[b,a,l,k]{A.~G.~Vieregg}
\author[f]{M.~R.~Vissers}
\author[c]{G.~Wang}
\author[aa]{W.~L.~K.~Wu}
\author[c]{V.~Yefremenko}
\author[gg]{E.~Yilmaz}
\author[c,b]{C.~Yu}
\author[a]{J.~Zivick}

\affil[a]{Department of Astronomy and Astrophysics, University of Chicago, 5640 South Ellis Avenue, Chicago, IL, 60637, USA}
\affil[b]{Kavli Institute for Cosmological Physics, University of Chicago, 5640 South Ellis Avenue, Chicago, IL, 60637, USA}
\affil[c]{High-Energy Physics Division, Argonne National Laboratory, 9700 South Cass Avenue, Lemont, IL, 60439, USA}
\affil[d]{Kavli Institute for Particle Astrophysics and Cosmology, Stanford University, 452 Lomita Mall, Stanford, CA, 94305, USA}
\affil[e]{SLAC National Accelerator Laboratory, 2575 Sand Hill Road, Menlo Park, CA, 94025, USA}
\affil[f]{Quantum Sensors Division, National Institute of Standards and Technology, 325 Broadway, Boulder, CO, 80305, USA}
\affil[g]{Pritzker School of Molecular Engineering, University of Chicago, 5640 S Ellis Avenue, Chicago, IL 606037, USA}
\affil[h]{Department of Physics and Astronomy, University of New Mexico, Albuquerque, NM, 87131, USA}
\affil[i]{School of Physics and Astronomy, Cardiff University, Cardiff CF24 3YB, United Kingdom}
\affil[j]{Fermi National Accelerator Laboratory, MS209, P.O. Box 500, Batavia, IL, 60510, USA}
\affil[k]{Department of Physics, University of Chicago, 5640 South Ellis Avenue, Chicago, IL, 60637, USA}
\affil[l]{Enrico Fermi Institute, University of Chicago, 5640 South Ellis Avenue, Chicago, IL, 60637, USA}
\affil[m]{Department of Physics, University of California, Berkeley, CA, 94720, USA}
\affil[n]{Department of Physics, University of Illinois Urbana-Champaign, 1110 West Green Street, Urbana, IL, 61801, USA}
\affil[o]{Department of Astronomy, University of Illinois Urbana-Champaign, 1002 West Green Street, Urbana, IL, 61801, USA}
\affil[p]{X-ray Science Division, Argonne National Laboratory, 9700 South Cass Avenue, Lemont, IL, 60439, USA}
\affil[q]{Department of Physics \& Astronomy, University of Pennsylvania, 209 S. 33rd Street, Philadelphia, PA 19064, USA}
\affil[r]{Sorbonne Universit\'e, CNRS, UMR 7095, Institut d'Astrophysique de Paris, 98 bis bd Arago, 75014 Paris, France}
\affil[s]{Department of Astrophysical and Planetary Sciences, University of Colorado, Boulder, CO, 80309, USA}
\affil[t]{Department of Physics, University of Colorado, Boulder, CO, 80309, USA}
\affil[u]{Department of Physics, Boston University, Boston, MA 02215, USA}
\affil[v]{Department of Physics, Case Western Reserve University, Cleveland, OH, 44106, USA}
\affil[w]{Theiss Research, La Jolla, CA, 92037, USA}
\affil[x]{t0.technology, 2200-300 Rue Leo-Pariseau, Montreal, Q.C., H2X 4B3, Canada}
\affil[y]{Materials Sciences Division, Argonne National Laboratory, 9700 South Cass Avenue, Lemont, IL, 60439, USA}
\affil[z]{Institute of Multidisciplinary Research for Advanced Materials, Tohoku University, Sendai, 980-8577, Japan}
\affil[aa]{California Institute of Technology, 1200 East California Boulevard, Pasadena, CA, 91125, USA}
\affil[bb]{Department of Physics \& Astronomy, University of California, One Shields Avenue, Davis, CA 95616, USA}
\affil[cc]{Center for AstroPhysical Surveys, National Center for Supercomputing Applications, Urbana, IL, 61801, USA}
\affil[dd]{School of Physics, University of Melbourne, Parkville, VIC 3010, Australia}
\affil[ee]{Department of Physics, Villanova University, 800 E Lancaster Ave., Villanova, PA 19085, USA}
\affil[ff]{Center for Astrophysics \textbar{} Harvard \& Smithsonian, 60 Garden Street, Cambridge, MA, 02138, USA}
\affil[gg]{Department of Electrical and Computer Engineering, University of Illinois Urbana-Champaign, 306 N Wright St, Urbana, IL 61801, USA}


%% file: acknowledgements.tex
H. Athreya's work is partially supported by the University of Chicago Physical Sciences Division's Eckhardt Graduate Scholarship.
The South Pole Telescope program is supported by the National Science Foundation (NSF) through awards OPP-2332483 and OPP-2408494.
Argonne National Laboratory's work was supported by the U.S. Department of Energy, Office of High Energy Physics, under contract DE-AC02-06CH11357.
Work performed at the Center for Nanoscale Materials, a U.S. Department of Energy Office of Science User Facility, was supported by the U.S. DOE, Office of Basic Energy Sciences, under Contract No. DE-AC02-06CH11357.
This document was prepared by the SPT-3G+ collaboration using the resources of the Fermi National Accelerator Laboratory (Fermilab), a U.S. Department of Energy, Office of Science, Office of High Energy Physics HEP User Facility. Fermilab is managed by Fermi Forward Discovery Group, LLC, acting under Contract No. 89243024CSC000002.
The SLAC group is supported in part by the Department of Energy at SLAC National Accelerator Laboratory, under contract DE-AC02-76SF00515.
This research was funded by the National Institute of Standards and Technology (ror.org/05xpvk416) and the University of Chicago (ror.org/024mw5h28) under agreement number 300004034.
This work is supported by the Gordon and Betty Moore Foundation, through Award \#14367.
The work at Case Western Reserve University is partially supported by the U.S. Department of Energy under award number DE-SC0009946.
Work at the University of Illinois Urbana-Champaign is partially supported by the U.S. Department of Energy under award number DE-SC0015655.
D.R.B. and W.D. were supported by DOE HEP under award DE-SC0021435.
The Melbourne authors acknowledge support from the Australian Research Council's Discovery Project scheme (No. DP210102386). 
The Paris group has received funding from the European Research Council (ERC) under the European Union's Horizon 2020 research and innovation program (grant agreement No 101001897), and funding from the Centre National d'Etudes Spatiales. 
R. A. Lew's work was performed under financial assistance award 70NANB21H182 and 70NANB25H091 from the National Institute of Standards and Technology, U.S. Department of Commerce. The statements, findings, conclusions, and recommendations are those of R. A. Lew and do not necessarily reflect the views of the National Institute of Standards and Technology or the U.S. Department of Commerce.